\documentclass[aps,amsfont,amssymb,floats,preprint,superscriptaddress,floatfix]{revtex4}
\usepackage{epsfig,graphicx}
\newcommand{\lnG}{\widetilde{\log}}
\newcommand{\expG}{\widetilde{\exp}}
\newcommand{\lambdaG}{\lambda}
\newcommand{\alphasens}{\alpha^{\mathrm{av}}_{\mathrm{sens}}}
\begin{document}
%\draft
%\preprint{}
%
\title{
Statistical descriptions of nonlinear systems at the onset of
chaos}

\author{Massimo Coraddu}
\email{massimo.coraddu@ca.infn.it} \affiliation{Dipart. di Fisica
dell'Universit\`a di Cagliari,
             I-09042 Monserrato, Italy}
\affiliation{Ist. Naz. Fisica Nucleare (I.N.F.N.) Cagliari,
             I-09042 Monserrato, Italy}
\author{Marcello Lissia}
\email{marcello.lissia@ca.infn.it} \affiliation{Ist. Naz. Fisica
Nucleare (I.N.F.N.) Cagliari,
             I-09042 Monserrato, Italy}
\affiliation{Dipart. di Fisica dell'Universit\`a di Cagliari,
             I-09042 Monserrato, Italy}
\author{Roberto Tonelli}
\email{roberto.tonelli@dsf.unica.it}
\affiliation{Dipart. di Fisica dell'Universit\`a di Cagliari,
             I-09042 Monserrato, Italy}
\affiliation{INFO-SLACS Laboratory, I-09042 Monserrato, Italy} 
\affiliation{Ist. Naz. Fisica Nucleare (I.N.F.N.) Cagliari,
             I-09042 Monserrato, Italy}
\date{November 28, 2005}
%\date{\today}
%
\begin{abstract}
Ensemble of initial conditions for nonlinear maps can be described
in terms of entropy. This ensemble entropy shows an asymptotic
linear growth with rate $K$. The rate $K$ matches the logarithm of
the corresponding asymptotic  sensitivity to initial conditions
$\lambda$. The statistical formalism and the equality $K=\lambda$
can be extended to weakly chaotic systems by suitable and
corresponding generalizations of the logarithm and of the entropy.
Using the logistic map as a test case we consider a wide class of
deformed statistical description which includes Tsallis, Abe and
Kaniadakis proposals. The physical criterion of finite-entropy
growth $K$ strongly restricts the suitable entropies. We study how
large is the region in parameter space where the generalized
description is useful.
\end{abstract}
 \maketitle
%%%%%%%%%%%%%%%%%%%%%%%%%%%%%%%%%%%%%%%%%%%%%%%%%%%%%%%%%%%%%%%%%%%%%%
\section{Introduction}
Sensitivity to initial conditions in chaotic systems has an
exponential asymptotic regime in fully-chaotic regions and a
power-law regime in transition regions. A description in terms of
generalized exponentials leading to the definition of generalized
Lyapunov exponents gives an unified framework for both
regimes~\cite{Tsallis:1997,Tonelli:2004ha}. For instance, if
Tsallis' generalization is used:
 the sensitivity $ \xi\equiv \lim_{\Delta x(0)\to 0} \Delta x(t)
/ \Delta x(0)$ grows asymptotically as the generalized exponential
$ \expG(\lambdaG t)$, where $\expG(x) = \exp_q(x) \equiv
[1+(1-q)x]^{1/(1-q)}$; the exponential behavior for the chaotic
regime is recovered for $q\to1$: $\lim_{q\to1} \exp_q(\lambda_q t)
= \exp(\lambda t)$. A large class of generalized exponentials 
shows similar behavior~\cite{Tonelli:2004ha}.

The statistical definition of entropy production rate, where an
ensemble of initial conditions confined to a small region is let
evolve and the entropy is a functional of the occupation numbers
of an appropriate partition, shows close analogy to the production
rate of thermal entropy and appears to coincide with the
Kolmogorov-Sinai entropy in chaotic regimes~\cite{Latora:1999prl}.

This rate of loss of information can be also suitably generalized
to include both full-chaotic and edge-of-chaos cases. At the edge
of chaos the statistical description is recovered with the
generalized entropic form proposed by
Tsallis~\cite{Tsallis:1987eu} $S_q = (1-\sum_{i=1}^{N}
p_i^q)/(q-1)$, which grows linearly for a specific value of the
entropic parameter $q$ characteristic of the system:
 $\lim_{t\to\infty}\lim_{L\to 0} S_q(t)/ t = K_q$, where $S_q $
reduces to $-\sum_{i=1}^{N} p_i\log p_i$ in the limit $q\to 1$,
$p_i$ being the fraction of the ensemble found in the $i$-th cell
of linear size $L$. As a matter of fact a large class of 
entropies, which includes Tsallis'one,  reproduces this 
asymptotic linear behavior~\cite{Tonelli:2004ha,Lissia:2005by}. 
The asymptotic power law behavior (exponent) that characterizes 
the entropy growth is the same that describes the asymptotic
 power-law sensitivity to initial
conditions \cite{Latora:1999vk,Tonelli:2004ha,Lissia:2005by}.

Finally, it has been also conjectured that the relationship
between the asymptotic entropy-production rate $K$ and the
Lyapunov exponent $\lambda$ for  chaotic systems (the Pesin-like
identity~\footnote{Pesin identity relates Kolmogorov-Sinai entropy
to the Lyapunov exponents; $K$ is in principle different.}:
$K=\lambda$) can be extended to systems at the edge of chaos $K_q 
= \lambda_q$~\cite{Tsallis:1997}.

 There exist numerical evidences supporting
this framework with the entropic form
 $S_q $ for the
logistic~\cite{Tsallis:1997} and generalized logistic-like
maps~\cite{Tsallis:1997cl}. The linear behavior of the generalized
entropy  for specific choices of $q$ has been observed for two
families of one-dimensional dissipative maps \cite{Tirnakli:2001}.

Evidence exists also for a more generalized coherent statistical 
framework, which includes Tsallis' 
proposal~\cite{Tonelli:2004ha,Lissia:2005by,Kaniadakis:2004td}. 
The generalized entropic form is $-\sum_{i=1}^{N} p_i\lnG (p_i)$, 
where $\lnG$ indicates the following two-parameter 
family~\cite{Mittal,Taneja1,Borges1,Kaniadakis:2004nx,Kaniadakis:2004ri} 
of logarithms
\begin{equation}\label{eq:logGen}
    \lnG(\xi) \equiv \frac{\xi^{\alpha}-\xi^{-\beta}}{\alpha+\beta}
    \quad ,
\end{equation}
where $\alpha$ ($\beta$) characterizes the large (small) argument
asymptotic behavior;  in particular, $\expG(\lambdaG t)\equiv
\lnG^{-1}(\lambdaG t)\sim (\lambdaG t)^{1/\alpha}$ for large $t$.
The requirements~\cite{Naudts1} that the logarithm be an
increasing function, therefore invertible, with negative concavity
(the related entropy is stable), and that the exponential be
normalizable for negative arguments  select $0 \leq \alpha\leq 1 $
and $0\leq \beta <1 $~\cite{Kaniadakis:2004td}; solutions with
negative $\alpha$ and $\beta$ are duplicate, since the family
(\ref{eq:logGen}) possesses the symmetry $\alpha \leftrightarrow
-\beta $.

Renormalization-group methods have been
used~\cite{Baldovin:2002a,Baldovin:2002b} to demonstrate that the
asymptotic behavior of the sensitivity to initial conditions in
the logistic and generalized logistic maps is bounded by a
specific power-law whose exponent could be determined analytically
for initial conditions on the attractor. For evolution on the
attractor the connection between entropy and sensitivity has been
studied analytically and numerically for Tsallis' entropic form\
\cite{Baldovin:2004}.

Sensitivity and entropy production have been studied in
one-dimensional dissipative maps using ensemble-averaged initial
conditions chosen uniformly over the interval embedding the
attractor~\cite{Ananos:2004a}: the statistical picture, i.e., the
relation between sensitivity and entropy exponents and the
generalized Pesin-like identity, has been confirmed. Indeed the
ensemble-averaged initial conditions appear more relevant for the
connection between ergodicity and chaos and for practical
experimental settings.

The main objective of the present work is to review the general
validity of the above-described picture, including the generalized
Pesin-like identity, and the physical criterion of finite-entropy
production per unit time (linear growth), which strongly selects
appropriate entropies and fixes their parameters. In addition we
specifically study how large is the transition region where the
generalized formalism is useful, before the usual exponential
regime sets in.

%%%%%%%%%%%%%%%%%%%%%%%%%%%%%%%%%%%%%%%%%%%%%%%%%%%%%%%%%%%%%%%%%%%%%%%%%%%%
\section{Statistical description}

From our investigation on the entire class of logarithms in
Eq.~(\ref{eq:logGen}), we review results for the following
one-parameter cases:

(1) Tsallis' seminal proposal~\cite{Tsallis:1987eu}: $\alpha=1-q$ 
and $\beta=0$;

(2) the $\gamma$ logarithm~\footnote{The exponentials 
corresponding to the $\gamma$, Tsallis, and Kaniadakis 
logarithms, $\expG(y)=\lnG^{-1}(y)$, are among the members of the 
class that can be explicitly inverted in terms of simple known 
functions~\cite{Kaniadakis:2004td}.}: $\alpha=2\beta=2\gamma$;

(3) the Abe logarithm:  $\beta=\alpha/(1+\alpha)=1-q_A$, named 
after the entropy introduced by Abe~\cite{Abe:1997qg}, which 
possesses the symmetry $q_A\to 1/q_A$;

(4) the Kaniadakis logarithm: $\alpha=\beta=\kappa$, which shares
the same symmetry group of the relativistic momentum
transformation and has applications in cosmic-ray and plasma
physics~\cite{Kaniadakis:2001nl,Kaniadakis:2002sr,Kaniadakis:2005zk}.

Our laboratory is the logistic map $x_{i+1}=1-\mu x^2_i$, at and
near the infinite bifurcation point $\mu_{\infty}=1.401155189$. If
the sensitivity  follows a deformed exponential $\xi(t) =
\expG(\lambdaG t)$, analogously to the chaotic regime when
$\xi(t)\sim \exp(\lambda t)$, the corresponding deformed logarithm
of $\xi$  should yield a straight line
$\lnG(\xi(t))=\lnG(\expG(\lambdaG t))=\lambdaG t$.

Starting from an initial condition $x_0$, the sensitivity has been
obtained, $\xi(t)= (2\mu)^{t}\prod_{i=0}^{t-1}|x_{i}|$, for $1
\leq t \leq 80 $; the  generalized logarithm $\lnG(\xi)$ has been
averaged over a sample of $4\times 10^{7}$ random initial
conditions $-1<x_0<1$. The averaging over initial conditions,
$\langle\cdots\rangle$, is appropriate for a comparison with the
entropy production.

For each of the generalized logarithms,
$\langle\lnG(\xi(t))\rangle$ has been fitted to a quadratic
function for $1\leq t \leq 80$ and $\alpha$ has been chosen such
that the coefficient of the quadratic term be zero; in fact
$\lnG(\xi)$ linear in $t$ means that the sensitivity $\xi$ behaves
as $\expG(\lambda t)$: we label this value $\alphasens$.

In fact, the exponent obtained with this procedure has been
denoted $q_{\mathrm{sens}}^{\mathrm{av}}$ in the case of Tsallis'
entropy~\cite{Ananos:2004a} and it is different from
$q_{\mathrm{sens}}$ obtained by choosing the initial condition
$x_0$ (fixed point of the map)~\cite{Baldovin:2004}.

The values of $\alphasens$ corresponding to the four different
choices of the logarithm are (1) 0.644, (2) 0.656, (3) 0.657, and
(4) 0.653 with a statistical errors of 0.002, calculated by
repeating the fitting procedure for sub-samples, and  a systematic
error of $0.004$, estimated  by fitting over different ranges of
$t$. The exponent of Tsallis' formulation $\alphasens = 0.644\pm
0.002$ is consistent with the value of Ref.~\cite{Ananos:2004a}
$q^{\mathrm{av}}_{\mathrm{sens}}=1-\alphasens\approx 0.36$. The
values of $\alphasens$ obtained using the four different
formulations are within $\pm1\%$: the common asymptotic behavior
(deviations at the 1\% level are due to the inclusion in the
global fitting of small values of $\xi$) is $\alphasens=0.650\pm
0.005$.

\begin{figure}[ht]
\includegraphics[width=\columnwidth,angle=0]{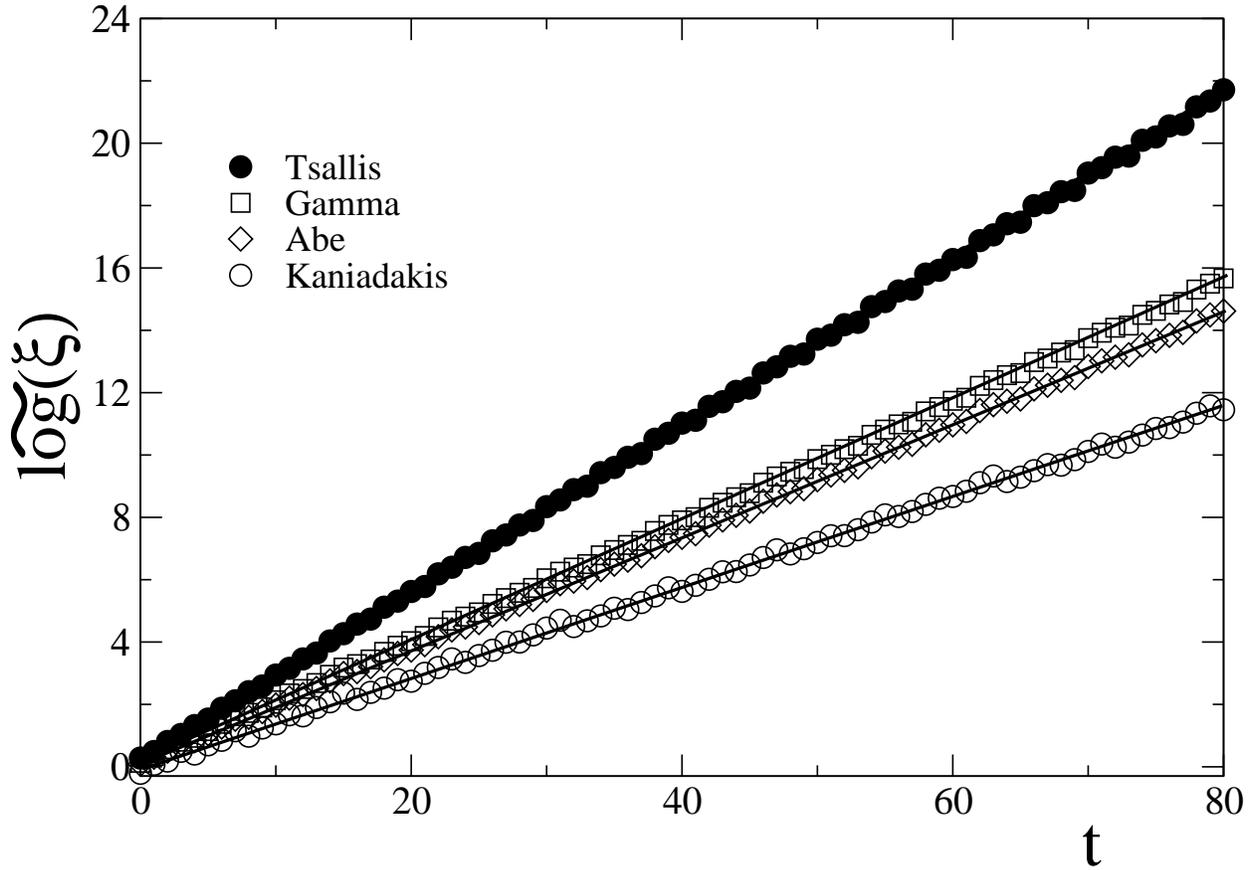}
\caption{Generalized logarithms $\lnG(\xi)$, see
Eq.~(\ref{eq:logGen}), of the sensitivity to initial conditions
averaged over $4\times 10^{7}$ uniformly distributed initial
conditions as function of time. From top to bottom: Tsallis',
$\gamma$, Abe's, and Kaniadakis' logarithms. The linear rising
behavior has been obtained with the asymptotic power $\alphasens$.
 \label{fig:sensitivity}}
\end{figure}

Figure~\ref{fig:sensitivity} shows the straight-line behavior of
$\lnG(\xi)$  for  $\alpha=\alphasens$:
 the corresponding slopes $\lambda$ (generalized Lyapunov
exponents) are (1) $0.271\pm 0.003$, (2) $0.198\pm 0.002$, (3)
$0.185\pm 0.002$, and (4) $0.148\pm 0.001$ from top to bottom; an
additional systematic error of about 0.003 has been estimated by
different choices of the range of $t$. While the values of
$\alpha$ are consistent with a universal exponent independent of
the particular deformation, the slope $\lambda$ strongly depends
on the choice of the logarithm.

The entropy has been calculated by dividing the interval $(-1,1)$
in $W=10^{5}$ equal-size boxes, putting at the initial time
$N=10^{6}$ copies of the system with a uniform random 
distribution within one box, and then letting the systems evolve 
according to the map. At each time $p_i(t)\equiv n_i(t)/N$, where 
$n_i(t)$ is the number of systems found in the $i$-th box at time 
$t$, the entropy of the ensemble is
\begin{equation}\label{eq:entropyGen}
    S(t) \equiv \left\langle \sum_{i=1}^{W} p_i(t) \lnG(\frac{1}{p_i(t)})\right\rangle =
    \left\langle \sum_{i=1}^{W} \frac{p_i^{1-\alpha}(t)-p_i^{1+\beta}(t)}{\alpha+\beta}
         \right\rangle
\end{equation}
where $\langle\cdots\rangle$ is an average over $2\times 10^{4}$
experiments, each one starting from one box randomly chosen among
the $W$ boxes. The choice of the entropic
form~(\ref{eq:entropyGen}) is fundamental for a coherent
statistical picture: the usual constrained variation of the
entropy in Eq.~(\ref{eq:entropyGen}) respect to $p_i$ yields as
distribution the deformed exponential $\expG(x)$ whose inverse is
indeed the logarithm appearing in
Eq.~(\ref{eq:logGen})~\cite{Kaniadakis:2004td}.

\begin{figure}[ht]
\includegraphics[width=0.8\columnwidth,angle=-90]{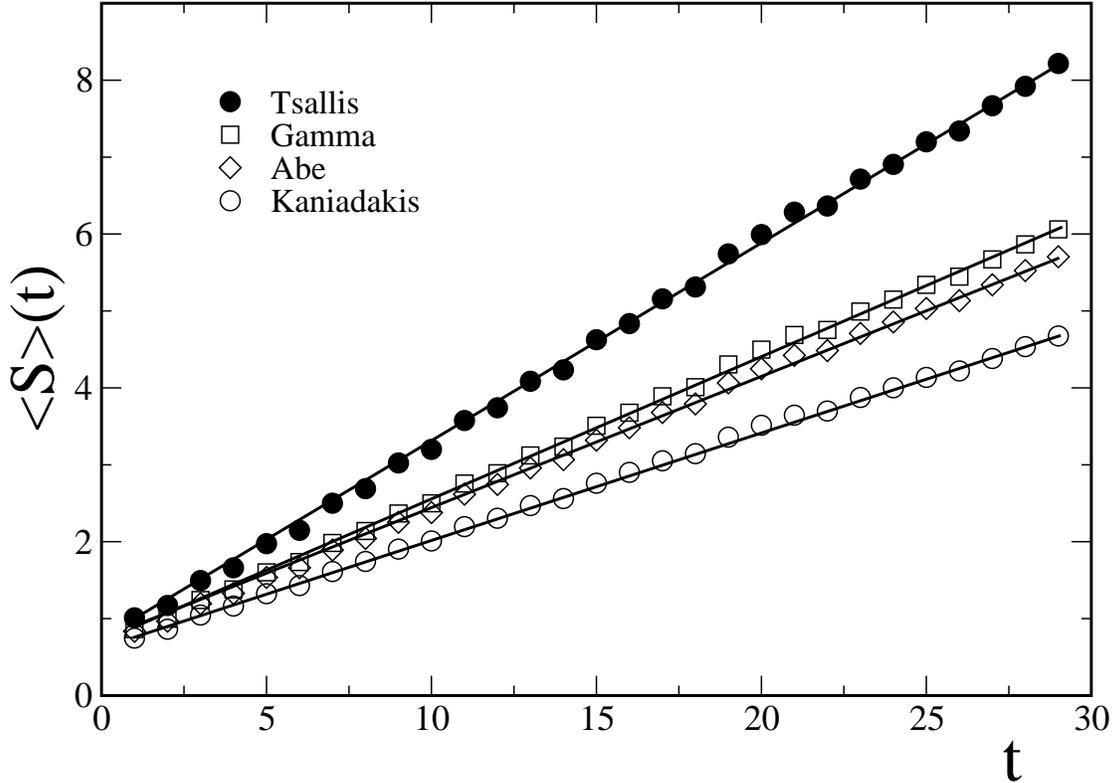}
\caption{Entropy as function of time, averaged over $2\times
10^{4}$ experiments, each one with an ensemble of $10^{6}$ points
over a grid of $10^5$ boxes. The entropies belong to the class of
Eq.~(\ref{eq:entropyGen}) and are defined with exactly the same
exponents $\alpha$ and $\beta$ used for the sensitivities; as in
Fig.~\ref{fig:sensitivity},  the curves show Tsallis', $\gamma$,
Abe's, and  Kaniadakis' entropies from top to bottom. Straight
lines are guides to the eyes. \label{fig:entropy} }
\end{figure}

Analogously to the strong chaotic case, where an exponential
sensitivity ($\alpha=\beta=0$) is associated to a linear rising
Shannon entropy, which is defined in terms of the usual  logarithm
($\alpha=\beta=0$), and consistently with the conjecture in
Ref.~\cite{Tsallis:1997}, the same values $\alpha$ and $\beta$ of
the sensitivity are used in Eq.~(\ref{eq:entropyGen}):
Fig.~\ref{fig:entropy} shows that this choice leads to entropies
that grow also linearly. This linear behavior is lost for values
of the exponent $\alpha$ different from $\alphasens$, confirming
for the whole class~(\ref{eq:entropyGen}) what was already known
for the $q$-logarithm~\cite{Tsallis:1997,Ananos:2004a}.

The resulting rate of growth of several entropies
$K_{\beta}=S_{\beta}(t)/t$ are (1) $0.267\pm0.003$, (2) $0.197.\pm
0.002$, (3) $0.186\pm 0.002$, and (4) $0.152\pm 0.002$, where the
statistical errors have been again estimated by sub-sampling the
experiments. While this rate $K$ depends on the choice of the
entropy, the generalized Pesin-like identity holds for each given
deformation:
\begin{equation}
K_{\beta}=\lambda_{\beta} \quad. \label{eq:pesin}
\end{equation}

\section{Width of the transition region}
Since we follow the system for a finite amount of time and the
sensitivity and entropy at each time step has a finite error, we
expect that the statistical description of the system using a
generalized formulation is practically useful not only at the
precise onset of chaos, but also for values of $\mu$ slightly
larger (fully chaotic region) or smaller (periodic region). To
estimate the width of the interval of $\mu$, we have repeated the
two previous experiments (sensitivity and entropy) with nine
values of $\mu$: four above and four below the onset of chaos.

\begin{figure}[ht]
\includegraphics[width=0.8\columnwidth]{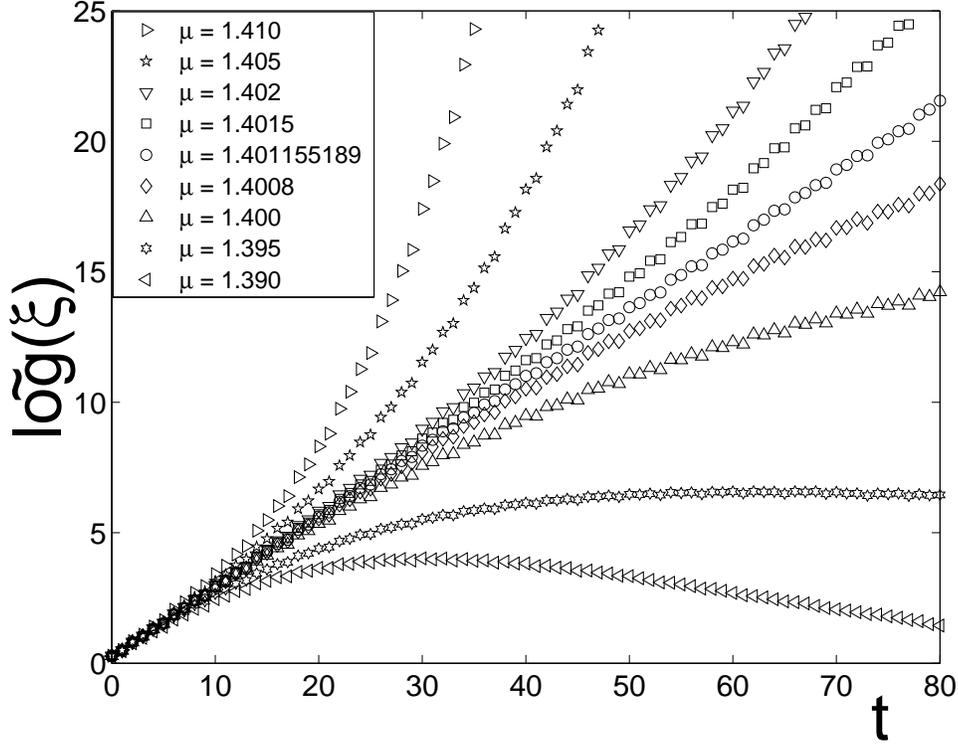}
\caption{Generalized logarithms $\lnG(\xi)$, see
Eq.~(\ref{eq:logGen}), of the sensitivity to initial conditions
averaged over $10^{7}$ uniformly distributed initial conditions 
as function of time in the case of Tsallis' formulation 
($\beta=0$) for nine values of $\mu$. The central curve 
($\mu=1.401155189$) is the top curve of 
Fig.~\ref{fig:sensitivity}; the four larger $\mu$ are in the 
chaotic region $\lambda>0$, the four smaller $\mu$ in the 
periodic one $\lambda<0$.\label{fig:sensitivityError} }
\end{figure}

\begin{figure}[ht]
\includegraphics[width=0.8\columnwidth]{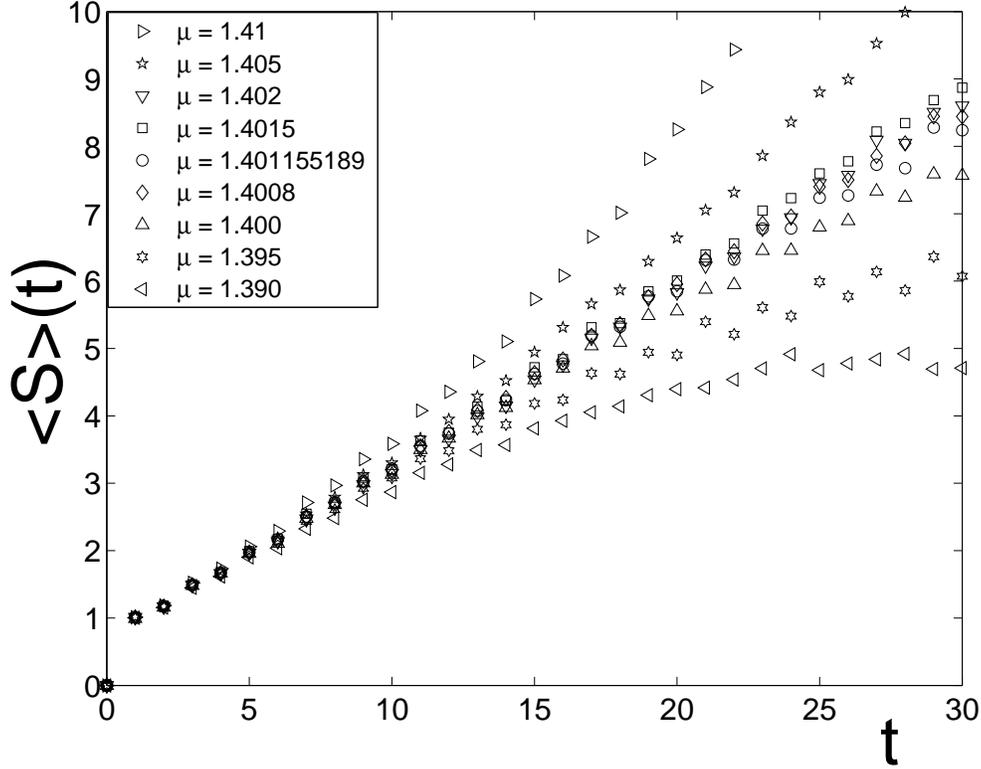}
\caption{Tsallis' entropy as function of time for the same nine
values of $\mu$ of Fig.~\ref{fig:sensitivityError}, averaged over
$10^{4}$ experiments, each one with an ensemble of $10^{6}$ 
points over a grid of $10^5$ boxes. \label{fig:entropyError} }
\end{figure}

In Fig.~\ref{fig:sensitivityError} we report our result for the
sensitivity in the case of Tsallis' formulation: analogous results
have been found for the other formulations. These curves should be
compared with the top curve in Fig.~\ref{fig:sensitivity}. It
appears that for $t< 30$ all systems with $1.400
\lessapprox\mu\lessapprox 1.402$ can be described by the
generalized formalism. This estimate is confirmed by the
corresponding curves for the entropy in
Fig.~\ref{fig:entropyError}.

For longer times the window shrinks: at $t=80$ already values of
$\mu$ different by $0.03 \%$ from the critical values deviate from
the power-law behavior (see Fig.~\ref{fig:sensitivityError}).

\section{Conclusions}

In summary, numerical evidence corroborates and extends Tsallis' 
conjecture that  also weak chaotic systems can be described by an 
appropriate statistical formalism. Such extended formalisms 
should verify precise requirements (concavity, Lesche 
stability~\cite{Kaniadakis:2004td,Kaniadakis:2003ls,Scarfone:2004ls}, 
and finite-entropy production per unit time) to both correctly 
describe chaotic systems and provide a coherent statistical 
framework: the last criterion restricts the entropic forms to the 
ones with the correct asymptotic behavior. Using  a specific 
two-parameter class that meets all these requirements, the 
logistic map shows:

(i) a power-low sensitivity to initial condition with a specific
exponent $\xi\sim t^{1/\alpha}$, where $\alpha = 0.650\pm 0.005$;
this sensitivity  can be described by deformed exponentials with
the same asymptotic behavior $\xi(t)=\expG(\lambdaG t)$ (see
Fig.~\ref{fig:sensitivity} for examples);

(ii) a constant asymptotic entropy production rate (see
Fig.~\ref{fig:entropy}) for trace-form entropies with a specific
power-behavior $p^{1-\alpha}$ in the limit of small probabilities
only when the exponent $\alpha$ is the same appearing in the
sensitivity;

(iii) a generalized Pesin-like identity holds $S_{\beta}/t\to
K_{\beta} = \lambda_{\beta}$ for each choice of entropy and
corresponding exponential in the class; the value of
$K_{\beta}=\lambda_{\beta}$ depends on the specific entropy and it
is not characteristic of the map.

We remark that the physical criterion of requiring that the
entropy production rate reach a finite and non-zero asymptotic
value has to consequences: (1) it  selects a specific value of the
parameter $\alpha$ ($\alpha$ is characteristic of the system); (2)
strongly restricts the kind of acceptable entropies to the ones
that have asymptotic power-law behavior (for instance it excludes
Renyi entropy~\cite{Johal:2004,Lissia:2005by}). The reason we ask
for a finite non-zero slope is that otherwise we would miss an
important characteristic of the system: its asymptotic exponent.

Finally we have estimated the range of $\mu$ for which the
sensitivity to initial conditions and the entropy are well
described by a power law behavior. The answer is clearly dependent
on the maximum time considered and on our experimental resolution.
For time smaller than 30 (80) the range appears of the order of
 0.07 \% (0.03\%)

%%%%%%%%%%%%%%%%%%%%%%%%%%%%%%%%%%%%%%%%%%%%%%%%%%%%%%
\begin{acknowledgments}

This work was partially supported by MIUR (Ministero
dell'Istruzione, dell'Universit\`a e della Ricerca) under
MIUR-PRIN-2003 project ``Theoretical Physics of the Nucleus and
the Many-Body Systems.''
\end{acknowledgments}
%%%%%%%%%%%%%%%%%%%%%%%%%%%%%%%%%%%%%%%%%%%%%%%%%%%%%%%%%%%%%%%%%%%%%

\end{document}